\documentstyle[11pt,newpasp,twoside,psfig]{article}
\begin{document}
\title{The Young Stellar Group Associated with HD 199143}
\author{Mario E. van den Ancker}
\affil{Harvard-Smithsonian Center for Astrophysics, 60 Garden Street, MS 42, 
Cambridge, MA 02138}
\author{Mario R. P\'erez}
\affil{Emergent-IT Corp., 9315 Largo Drive West, Suite 250, Largo  MD 20774}
\author{Dolf de Winter}
\affil{TNO-TPD, Stieltjesweg 1, 2600 AD  Delft, The Netherlands}

\begin{abstract}
Recently, several groups of young stars in the solar neighborhood 
have been discovered.  Given their proximity, these systems are 
ideally suited for detailed studies of star and planet formation.  
Here we report on a group of young stars associated with the 
bright F8V star HD~199143.  
At a distance of only 48~pc, this is the closest YSO group 
containing a classical T~Tauri star (HD~358623; K7--M0e).  
New ground-based mid-infrared data shows that both HD~199143 
and HD~358623 have large infrared excesses due to 
circumstellar disks.  A systematic search for new 
members of this {\it Capricornius association} has yielded 
four new probable members, which we use to derive an 
age of 5--10 Myr for the group as a whole.
\end{abstract}

\section{Introduction}
In recent years, a fascinating picture of the recent star 
formation history of the solar neighborhood has emerged: 
10--40 million years ago an ensemble of molecular clouds 
were forming stars at a modest rate near the present position 
of the Sun. About 10~Myrs ago, the most massive of these 
newly formed stars exploded as a supernova, terminating the 
star formation episode and generating the very low density 
region seen in most directions from the present Sun. This 
scenario can not only explain the presence of young 
stellar groups close to the earth such as the TW Hydrae 
and the newly identified Tucanae Association (Kastner et al. 
1997; Zuckerman \& Webb 2000 and these proceedings), 
but also explains how the $\beta$~Pic moving group can be 
so young (20~Myr; Barrado y Navascu\'es et al. 1999), 
and yet so close.
\begin{figure}
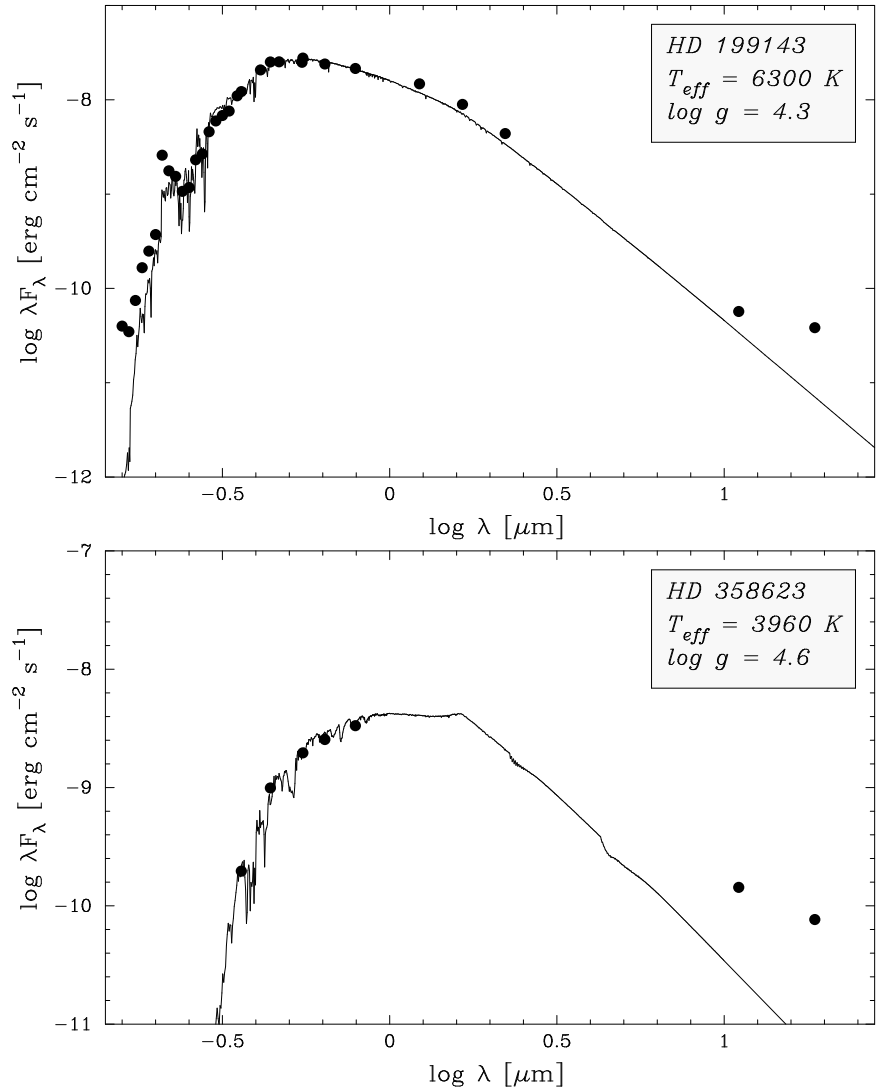

\centerline{\psfig{figure=sed_hd199143.ps,width=11.5cm,angle=270}}
\centerline{\psfig{figure=sed_hd358623.ps,width=11.5cm,angle=270}}
\caption[]{Spectral Energy Distributions of HD 199143 (top) 
and HD 358623 (bottom).  Also shown are Kurucz models for the 
stellar photospheres, fitted to the observed energy distribution.  
Note the presence of strong infrared excesses in both sources 
and the presence of excess UV emission in HD~199143.}
\end{figure}

In a recent {\it A\&A letter} (van den Ancker et al. 2000) 
we have identified two nearby objects, the bright F-type star 
HD~199143 and the late-type emission-line star HD~358623, 
as young stars.  In these {\it proceedings} we will 
extend this work by presenting new mid-infrared observations 
of HD~199143 and HD~358623, demonstrating that both possess 
circumstellar disks.  We will also present the results of 
a systematic search for further candidates, showing 
that both stars belong to a larger group of young stars, 
which we tentatively name the {\it Capricornius 
association}.  In the final section of this contribution 
we will discuss its relation to the other 
young stellar groups discussed in these proceedings and 
briefly touch upon its formation history.

\section{HD~199143 and HD~356823}
HD~199143 is a bright ($V$ = 7.27), nearby (Hipparcos distance 
of 47.7 $\pm$ 2.4~pc) F8V star, which would be completely 
inconspicuous if it hadn't been detected as a bright 
extreme-ultraviolet source by the {\it ROSAT} and {\it EUVE} 
missions.  A recent study of the optical and UV spectrum of 
HD~199143 by van den Ancker et al. (2000) revealed the presence 
of emission lines of Mg\,{\sc ii}, C\,{\sc i}, C\,{\sc ii}, 
C\,{\sc iii}, C\,{\sc iv}, Si\,{\sc iv}, He\,{\sc ii} and 
N\,{\sc v} and a large amount variability, both in the continuum 
and line fluxes.  The fact that these phenomena were only 
found in the ultraviolet part of the spectrum suggests 
that HD~199143 is a 
binary system, consisting of a rapidly rotating F-type 
primary and a low-mass chromospherically active companion 
which dominates the ultraviolet and infrared light of the 
system.

A literature search for sources near HD~199143 revealed that 
a photometrically variable K7--M0e dwarf, HD~358623 
(BD$-$17$^\circ$6128), is located only a few arcminutes 
from HD~199143.  This star was previously studied by Mathioudakis 
et al. (1995), who found strong H$\alpha$ emission and 
evidence for a high Li abundance, i.e. the characteristics 
of a classical T~Tauri star.  Data from the Tycho-2 catalog 
shows that HD~199143 and HD~358623 have identical proper 
motions.  Both the closeness of the two stars and the 
similarity of the space motions strongly suggest that the 
two stars must form a physical group.  The only explanation 
for the presence of two active stars of such different 
masses in such close proximity is to pose that the two 
stars are young.
\begin{figure}[t]
\centerline{\psfig{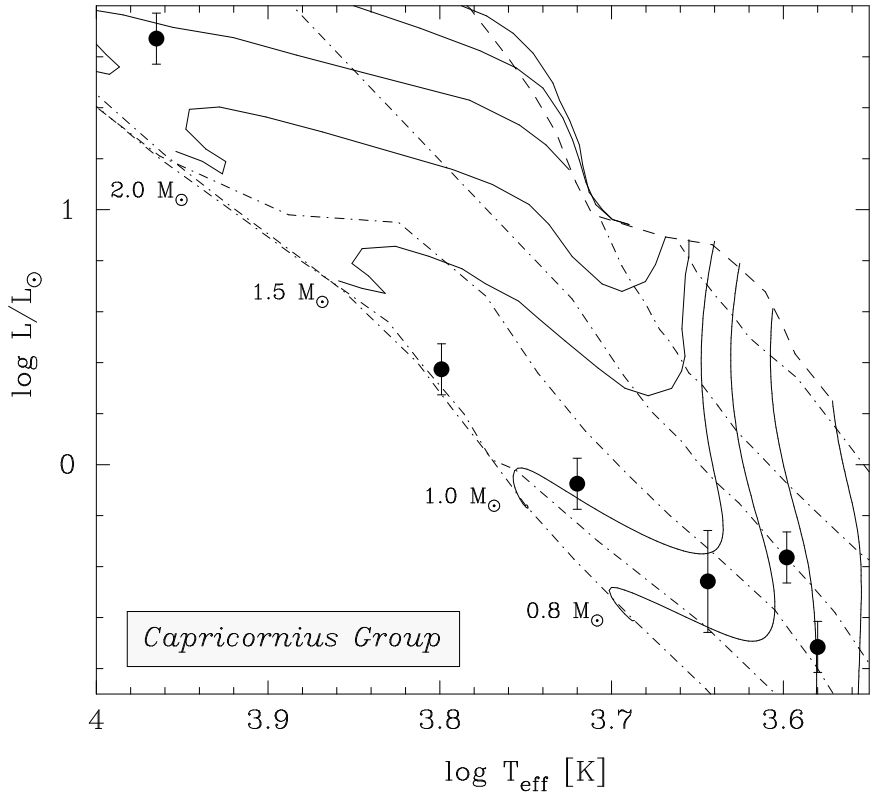}}
\caption[]{Hertzsprung-Russell diagram of the Capricornius group.  
Also shown are the evolutionary tracks (solid lines) and isochrones 
(dash-dotted) by Palla \& Stahler (1993).}
\end{figure}

New observations of HD~199143 and HD~356823 were obtained 
with the TIMMI2 instrument on the ESO 3.6m telescope.  Both 
sources were detected in the $N$ (11~$\mu$m) and $Q$ (19~$\mu$m) 
bands.  Spectral Energy Distributions (SEDs) of HD~199143 and 
HD~358623, which include the newly determined $N$ and 
$Q$ band data, are shown in Fig.~1.  As can be seen clearly 
from this figure, both stars possess large excesses above 
photospheric levels in the mid-infrared.  We explain these 
infrared excesses as being due to the presence of 
circumstellar disks in both systems.  In the case of 
HD~199143 it is not clear whether this disk is in 
the form of a disk around the low-mass companion, or 
is in the form of a circumbinary disk.

\section{Further Capricornius Association Members}
Based on a systematic search for stars near HD~199143 with 
strong X-ray fluxes we have identified four probable new 
members of our newly discovered association in Capricornius.  
All four have {\it ROSAT} Point Source Catalog fluxes which 
are an order of magnitude higher than those of normal late-type 
stars at 48~pc and have a projected distance of less than five 
degrees (4.2~pc at $d$ = 48~pc) from HD~199143.  Similarly 
strong X-ray sources are not found in adjacent fields of 
identical size, leading us to believe that these stars 
are indeed members of the Capricornius association.  For two 
of the newly identified candidate members, proper motions 
are available, which are in agreement with those 
of HD~199143 and HD~358623.

Further indications for the hypothesis that our four 
new candidate members belong to the Capricornius association 
comes from the observation that if we assume a common 
distance of 48~pc, HD~199143 and HD~358623, as well as the 
four new candidate members form a smooth curve in the 
Hertzsprung-Russell diagram (Fig.~2).  According to the 
models by Palla \& Stahler (1993), all stars are located 
between the isochrones with ages between $5 \times 10^6$ 
and $10^7$ years.  Although the youth of our newly selected 
candidate members remains 
to be confirmed through spectroscopic means, we conclude 
that most likely these four stars are indeed members of 
a more extended group of young stars associated with 
HD~199143.  Since the search we performed here for new 
members is certainly not complete (it is based on the 
catalog data in the Simbad database), we expect to 
be able to find additional low-mass members of the 
Capricornius association through dedicated imaging.  
Such observations are currently planned.

\section{Discussion and Conclusions}
In these proceedings we have argued that HD~199143 
and HD~358623 are part of a larger group of young 
stars, which we tentatively name the Capricornius
association.  Several of the properties of this 
newly discovered group make it unique: not only is it 
at a distance of 48~pc the closest association containing 
a {\it bona fide} classical T~Tauri star (the TW~Hya group 
is at slightly more than 50~pc), but its declination of 
$-17^\circ$ also clearly separates it from the other 
newly discovered YSO groups, which are all located 
much further to the South.

Yet kinematically our Capricornius association may 
be related to the other nearby regions of recent star 
formation.  HD~199143 has a galactic space velocity 
$(U, V, W)$ of $(-10 \pm 13, -13 \pm 6, -13 \pm 6)$ km~s$^{-1}$, 
similar to that found for the Tucanae and TW Hydra associations 
(Zuckerman \& Webb 2000).   This, as well as the similarity 
in the ages (5--10 Myr for Capricornius vs. $\sim$ 10 Myr 
for both Tuc and TW Hya), suggests that all three associations 
may have formed from the same cloud complex.  Star formation 
in this large cloud may have progressed linearly, as 
is also observed commonly in more distant star forming 
regions (Elmegreen et al. 2000 and references therein), 
starting with the most southern association (Horlogium, 
Tucanae, TW Hydra) and progressing to our newly 
identified Capricornius group.  Whether star formation 
ended here or whether more northern associations, and 
possibly even remnants of the parent molecular cloud, 
exist and remain to be discovered is a question 
that awaits further investigation.  In any case 
the Capricornius group represents a unique 
opportunity to not only gain 
a better understanding of the star formation history in 
the solar neighborhood, but to also allow us more 
insight in the structure of protoplanetary disks and 
hence our own origins.


\begin{references}
\reference
Barrado y Navascu\'es, D., Stauffer, J.R., Song, I., \& Caillault, J.P. 1999, ApJ 520, L123
\reference
Elmegreen, B.G., Efremov, Y., Pudritz, R.E., \& Zinnecker, H. 2000, 
 in Protostars \& Planets IV, ed. V. Mannings, A.P. Boss \& S.S. Russell 
 (Tucson: Univ. of Arizona Press), 179
\reference
Kastner, J.H., Zuckerman, B., Weintraub, D.A., \& Forveille, T. 1997, 
 Science 277, 67
\reference
Mathioudakis, M., Drake, J.J., Craig, N., et al. 
 1995, A\&A 302, 422
\reference
Palla, F., \& Stahler, S.W. 1993, ApJ 418, 414
\reference
van den Ancker, M.E., P\'erez, M.R., de Winter, D., \& McCollum, B. 2000, 
 A\&A 363, L25
\reference
Zuckerman, B., \& Webb, R.A. 2000, ApJ 535, 959
\end{references}
\end{document}